\documentclass[journal]{IEEEtran}
\ifCLASSINFOpdf
\else
\fi
\usepackage[numbers]{natbib}
\usepackage{epsfig}
\usepackage{epstopdf}
\usepackage[T1]{fontenc}
\usepackage{amssymb}
\usepackage{amsmath}
\usepackage{amsfonts}
\usepackage{verbatim}
\usepackage{graphicx}
\usepackage{hyperref}
\usepackage[usenames,dvipsnames]{xcolor}
  \usepackage[ruled]{algorithm}
\usepackage{algpseudocode}
\usepackage{array}
\usepackage{multirow}
\usepackage{caption}
\usepackage{subcaption}
\usepackage{multicol}
\usepackage{pgfplots}
\usepackage{lipsum}
\usepackage{colortbl}
\usepackage{soul}
\begin{document}
%
\title{IoT based Smart Access Controlled Secure Smart City Architecture Using Blockchain}

\author{\IEEEauthorblockN{Rourab Paul$^1$, Nimisha Ghosh$^2$, Suman Sau$^2$ Amlan Chakrabarti$^3$, Prasant Mahapatra$^4$\\}
\IEEEauthorblockA{Computer Science \& Engineering$^{1}$,Computer Science \& IT$^{2}$,  Siksha 'O' Anusandhan 
University, Odisha, India$^{12}$;  School of IT, University of Calcutta India$^3$;
Department of Computer Science University of California, USA$^4$,\\ mail \{rourabpaul, nimishaghosh, sumansau\}@soa.ac.in$^{12}$, acakcs@caluniv.ac.in$^3$, pmohapatra@ucdavis.edu$^4$}
}

%


\maketitle

\begin{abstract}

Standard security protocols like SSL, TLS, IPSec etc. have high memory and processor consumption which makes all these security protocols unsuitable for resource constrained platforms such as Internet of Things (IoT). Blockchain (BC) finds its efficient application in IoT platform to preserve the five basic cryptographic primitives, such as confidentiality, authenticity, integrity, availability and non-repudiation. Conventional adoption of BC in IoT platform causes high energy consumption, delay and computational overhead which are not appropriate for various resource constrained IoT devices. This work proposes a machine learning (ML) based smart access control framework in a public and a private BC for a smart city application which makes it more efficient as compared to the existing IoT applications. The proposed IoT based smart city architecture adopts BC technology for preserving all the cryptographic security and privacy issues. Moreover, BC has very minimal overhead on IoT platform as well. This work investigates the existing threat models and critical access control issues which handle multiple permissions of various nodes and detects relevant inconsistencies to notify the corresponding nodes. Comparison in terms of all security issues with existing literature shows that the proposed architecture is competitively efficient in terms of security access control.
\end{abstract}

%
\IEEEpeerreviewmaketitle

\section{Introduction}

The rapid growth of communication technology, network technology and increasing number of smart devices makes IoT very relevant for research exploration. Conversely the existing literature survey notices that IoT platform still suffers from privacy and security vulnerabilities \cite{jha}. The centralized architecture and less resource availability in most of the IoT devices have  made the conventional security and privacy preservation approaches inappropriate for IoT platforms \cite{sicari}, \cite{attackiot} \cite{das}. The  decentralized security and privacy on IoT applications can be facilitated by blockchain technology but conventional blockchain approaches have significant energy consumption, latency and execution overhead which are improper for most resource budgeted IoT platforms. 
\par This research exploration studies how the blockchain becomes lightweight and suitable for IoT based smart city applications \cite{future} with smart ML based access control. 
The authors in \cite{skarmeta} reported a decentralized capability based access control approach to
control access of sensitive information. This proposed method has high latency and overhead and also compromised with user privacy. In \cite{gross}, the authors used TLS and IPSec protocols for sensor data authentication and privacy but these approaches are very resource expensive for IoT platform. In \cite{ukil}, Ukil et. al reported a privacy management system which quantifies the risk of disclosing data to others. However, in many situations, the perceived benefit of IoT applications made the risk of privacy loss more significant. Hence it finds an obvious requirement of privacy-aware sharing of IoT sensor/output data without compromising the privacy of users. Though the works \cite{skarmeta} - \cite{ukil} talked about decentralization and privacy issues but they did not adopt blockchain and efficient access control management in their architecture.
The works in \cite{liu}, \cite{bhat}, \cite{frey} and there are also a few more works which were conducted in early years where authors stated about the different types of efficient access control mechanisms but they did not use full-fledged blockchain technology in their IoT platform. Liu et. al \cite{liu} proposed a role based access control system where roles indicate administrator and guest which are subjected to entities that access various resources within an organization. The associated roles with different access rights like read, write, execute are granted to IoT nodes. This role based access control schemes establish a many-to-many relations between the access rights and the nodes. In \cite{bhat}, the authors proposed an access control which is based on policies and it combines different types of attributes, like node attributes, objects like nodes that holds resources attributes etc. All these attributes set certain conditions for access rights grants to nodes. Both \cite{liu} and \cite{bhat} validate access rights of nodes that are usually performed by a centralized entity which implies an issue of single point of failure.
Frey et al. \cite{frey} have proposed capability based access control method where the access permission validation
is conducted by the requested IoT nodes instead of a centralized HP entity but IoT nodes usually have low resources. Hence it may be easily compromised by attackers. 
The authors in \cite{alidori}, \cite{zhang}, \cite{afar} have implemented smart contract using blockchain technology to achieve distributed trust and privacy of IoT platform.
Additionally blockchain utilizes resources of all participating nodes to address scalability and robustness which decrease many-to-one traffic flows. As a result it solves the issue of single point failure and delay issues. The inherent anonymity of blockchain is appropriate for IoT applications where the identity of the users is kept private. Blockchain technology serves a trustful network over dishonest nodes which is appropriate in IoT platforms where huge number of heterogeneous devices are interconnected. On the other-hand, straight forward adoption of blockchain in IoT is not possible due to certain constraints of IoT architecture. 
%

Alidori et. al \cite{alidori} proposed an access control feature in an IoT platform with cloud storage, service providers, user devices, local storage and smart homes where each consists of a miner and multiple IoT nodes. Each home controls a private local blockchain with a policy header. The policy header controls all the access requests related to the home. It has been observed that \cite{alidori} serves distributed and immutable storage for access control policies which wasted the capability of the blockchain.
The authors of \cite{zhang} proposed a smart contract-based access control framework, which has multiple access control contracts, one register contract and one judge contract. This framework was proposed to attain distributed and trustworthy access control for IoT applications. This proposal talks about access control algorithm which grants the permission of subject (sender) and object(receiver) by checking its access control header and charges a blocking time as a penalty if any time related inconsistency is detected. We have noticed several cases where time inconsistency did not occur due to attack of malicious node. Time inconsistency may also take place due to network congestion or for several other reasons. In \cite{zhang}, these false positive cases may charge unnecessary penalty. Authors of \cite{zhang} also overlooked other inconsistencies which may significantly affect IoT platform. 
Article \cite{afar} proposes an access control, where the blockchain plays the key role
of a decentralized access control management. The \cite{afar} used a tokens to represent access rights. The sender pads access control permissions into the locked scripts of the transaction. The
receiver unlocks the locking scripts to prove the possession of the token. This scheme granted node's access by receiving a token, and it grant access rights to another
node by delivering a token. The expensive computing requirement of locking scripts made the inefficient for access control management.
\par To address the limitations of the aforementioned works, this work proposes a smart city architecture which uses blockchain technology for distributed trust and privacy. The architecture also adopts a smart and efficient access control technique which considers all realistic inconsistencies of IoT platforms.  

This work proposes three main parts, namely smart block, canopy and storage. A city can be partitioned into several smart blocks.  All Smart blocks are equipped with sensors and outputs to collect and distribute real-time data according to the policy stated in access control header. The canopy network consists of smart blocks such as admin miners, administration authorities such as local police station, municipalities etc. The cloud storage is required to share data between smart blocks and others central/states administration etc. The transactions of different types of blockchains take place depending on its network hierarchy. 
The merit of this research proposal is to adopt blockchain based IoT architecture with efficient access control management which is lightweight in terms of computation and resource usage to delivers decentralized security and privacy. 
The main contributions of this proposal are given as follows:
\begin{itemize}
\item The proposed IoT based blockchain platform has two level of network hierarchy where the $1^{st}$ hierarchy level consists LP nodes (Local Network) and its admin named as $block~admin$ and the $2^{nd}$ hierarchy level (canopy network) consists of $block~admin$ and other high processing computers belonging to either same or different group. Both the hierarchies can handle all the privacy and security threats. The adoption of lightweight security such as symmetric key cryptography for the $1^{st}$ hierarchy level makes the $1^{st}$ hierarchy more efficient in terms of latency and resource budget. The $1^{st}$ and $2^{nd}$ hierarchy level blocks also achieve a fundamental security features such as confidentiality, integrity and availability, authenticity and non repudiation.
\item The proposed smart access control header controls all the honest transactions by restricting the malicious read write transaction requests. Any malicious activity termed as an inconsistency can be detected as data, network and time inconsistency. For data inconsistency detection, we have used Dempster-Shafer Theory of Evidence machine learning algorithm.
\item The automated code generation for LP node and HP node make the system user friendly. The repeated LP code download process using Over The Air (OTA) \cite{ota} protocol prevents execution of malicious LP node code for long term scenario. $Block~admin$ generates hash of the code of LP nodes to detect code alteration attack. 
\item The proposed IoT architecture is also appropriate for smart agriculture and smart home applications which can result in fast, secured and efficient public governance system .
\end{itemize}
The organization of the article is stated below. Sec. II states the
architecture of the proposed system. The work flow of the
architecture is described in sec. III. Auto code generator algorithm and  smart contract framework are stated in sec. IV \& sec. V respectively. Result, implementation and conclusion are organized in sec. V and VI respectively
\vspace{-10pt}
\section{Architecture}
\label{sec:arch}
The proposed architecture is divided into 3 different parts, such as smart block, canopy network and storage. The description of three parts are stated below.
\subsection{Smart Block}
\label{smrt:blk}
A city is partitioned into small blocks which are labelled as $smart~block$. Each $smart~block$ owns variety of sensors like camera, weather station sensor,  thermostat, health structure sensor, LDR etc as mentioned in table \ref{table:senslist}.
\begin{table}[!htb]
\vspace{-5pt}
\caption{Services in Proposed Smart City} 
\centering  
\resizebox{8cm}{!}{%
    \begin{tabular}{|c|c|c|c|}
        \hline
Service 	& Communication & Tolerable BC & Remarks\\
Name        & Rate/node     & Latency      & \\\hline
GPS         & 1 pkt per     & 30 min       & The node location is verified \\
            & 30 min        &              & in each 30 minutes\\\hline
Health      & 1 pkt per     & 30 min       & To check health condition \\
Structure   & 10 min        & 			   & of bridge and multi-storage\\\hline  
Weather     & On  	        & 1 min        & It measures temperature,  \\
Station     & Demand        & 			   & humidity and pressure\\\hline  
Air         & 1 pkt per     & 5 min        & green house sensors  \\
Quality     & 30 min        & 			   & will be used\\\hline  
Smart       & On    	    & 1 min        & Lights through out  \\
Light       & Demand        & 			   & city can be automated\\\hline       
Camera      & On     		& On       	   & e.g. the number of\\
~		    & demand        & Demand	   & can be counted\\\hline
Traffic     & 1 pkt per     & On       	   & The counted car adjust\\
Light		& 30 min        & Demand	   & time slot of traffic light\\\hline 
    \end{tabular}
}

\label{table:senslist} 
\end{table}

 All the mentioned sensors and outputs are installed in LP nodes and the $block~admin$ has lawful access to all sensor and output data of these smart LP nodes. A private blockchain along with an optional local storage is maintained by $block~admin$ to store various informations generated from LP nodes. Unlike the bitcoin's blockchain whose control is decentralized, here the local BC is managed centrally by its $block~admin$. All the transactions from or to the nodes are chained together by the $block~admin$. The $block~admin$ is solely responsible for adding new LP nodes or removing an existing LP node. The addition of new LP node is similar to 'create coin' transaction in bitcoin \cite{bitcoin}. There is an access control header owned by $block~admin$ to control all the transactions occurring in its block. Shared key using the Diffie-Hellman key exchange algorithm \cite{hans} permits all the transactions in this platform. 
 The proposed private blockchain used in this architecture avoids Proof of Work (PoW) to reduce the associated overheads. The nodes pad a pointer to the previous blocked copies of the policy in the previous block header to the next new block and chains the block to the blockchain. Unlike the bitcoin technology all the transactions are treated as honest transactions, whether the block is mined or not. The private blockchain is designed not only for user authentication purpose but also for mutual authentication between nodes, generating and securely storing operation details and outline-based IoT contracts. 
\subsection{Canopy Network}
This network is a peer to peer network which accommodates smart HP blocks like local police stations or state public administrative bodies. HP nodes in the canopy network are grouped in clusters and each group elects a Group Head (GH) to reduce the network overhead. Each GH controls a public blockchain. The GHs owns the list of public key of requesters who are permitted to access data for the smart blocks connected with this group. The GHs maintain public keys of respondents of nodes connected to that group which is allowed to be accessed. It is to be noted that the $block~admin$ control its own private blockchain and $block~admin$ is also a node of public blockchain mounted in canopy network.
\subsection{Cloud}
\begin{figure*}[!htb]
\centering
\includegraphics[scale=0.2]{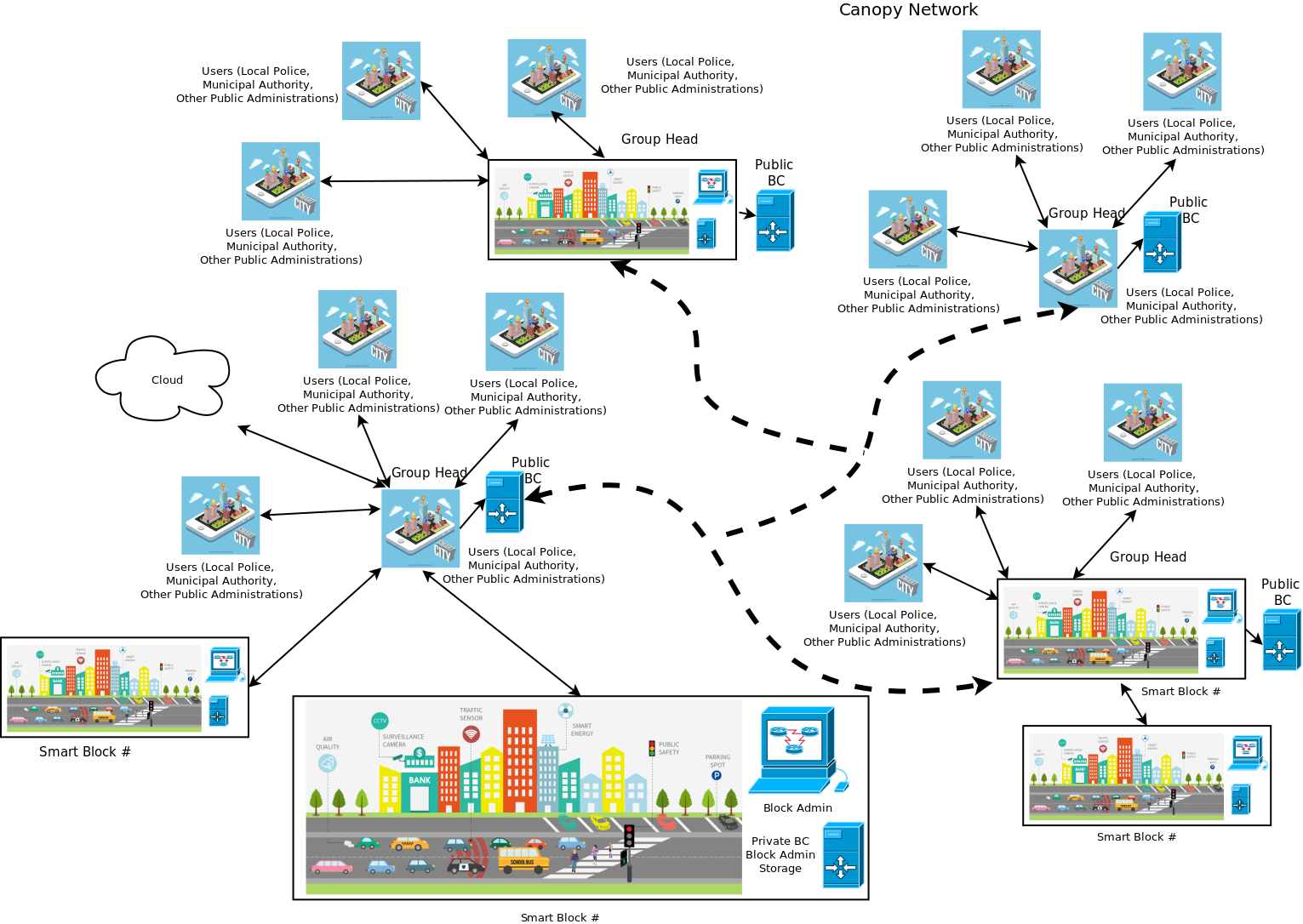}
\vspace{-6pt}
\caption{Architecture of Smart City}
\vspace{-8pt}
\label{fig:arch}
\end{figure*}

The cloud may also be a member of GH. In some applications, LP nodes of the smart block may want to store its data in the cloud storage. Those data should be accessed by the third party to provide certain services to the LP nodes. For example, in the federal structure of a country like India, few other state organizations or few central organizations want to access the data of LP nodes, they can read or in certain cases can write those through cloud.
All the transactions in LP nodes and canopy nodes are tagged as transactions. Four types of transactions may occur in this proposed platform. If the LP nodes of a smart block store data in the local storage of $block~admin$ or in the cloud, it is termed as write transaction. The read transaction will be coined if other states/central organization or $block~admin$ from same group or different group want to monitor cloud data or LP node data. Addition of a new node to the smart block is done by a \textbf{genesis} transaction and a device is removed by a \textbf{remove} transaction. It is to be noted that all the transactions to or from the smart block will be stored in the local blockchain. All the above transactions use shared key to encrypt their data. The data integrity of data is managed by lightweight hashing technique \cite{lwhash}. On a simpler note, it can be stated that the whole architecture is partitioned into two hierarchies such as $1^{st}$ level network hierarchy where private blockchain is running and $2^{nd}$ level network hierarchy where public blockchain is functional. The $1^{st}$ hierarchy consists of LP-HP nodes and the $2^{nd}$ hierarchy consists of HP-HP nodes. As shown in figure \ref{fig:archbc2}:a, the interfaces of these two hierarchies are done by $Block~Admin$. All the transactions of these two hierarchies are mostly identical. The pictorial representation of proposed smart city architecture including smart block, canopy network and cloud storage are detailed in fig. \ref{fig:arch}. 

\section{Work Flow of Architecture}
\label{sec:soa}
The work flow of the proposed architecture consists of two main parts. In initialization process, the pre-requisitions of LP platform (LP) is configured. The tables of databases of HP node such as $block~admin$ and $group~head$ are configured in initialization process. In the transaction process, it can be seen how the actual data is moved through the local and canopy network.
\vspace{-10pt} 
\subsection{Initialization}
The proposed model is involved with 2 types of initialization processes; intitialization of low processing (LP) platform and  initialization of high processing (HP) nodes in $block~admin$ or in canopy network.
\subsubsection{Low Processing (LP) Initialization}
\label{sec:lpinit}
Let us assume that $n_1$ sensors and $n_2$ output interfaces are connected with each LP node and total number of LP nodes under one $block~admin$ is $m$. Each LP node can collect data from $n_1$ number of sensors and can feed data to $n_2$ number of outputs. Each LP node has one device id or node id and three keys such as public key, private key and a key for symmetric key encryption. The LP nodes and $block~admin$ share light weight symmetric key \cite{lwhash}. The hash of public key represents node id of LP nodes which is very similar to the bitcoin platform where account id is also generated from hash value of its public key. The property of hash assured that hash of any random number is also unique \cite{arvind}. 
\subsubsection{High Processing (HP) Initialization}
The collected data from $n_1$ number of sensors are encrypted by each LP node and forwarded to the HP nodes using light weight cryptography algorithm \cite{lwhash}. The HP nodes accept encrypted data and check sender LP node's id, list of public keys, sensor id, transaction type, access control header, time stamps and hash values. If all of these checking processes are successful, the transaction will be accepted and copy of the transaction is stored in the private blockchain of $block~admin$ which may synchronized to public blockchain later. Two types of HP nodes are proposed in this report such as $block~admin$ and users in 1$^{st}$ network layer hierarchy and $group~head$ in canopy network or 2$^{nd}$ network layer hierarchy. All the transactions are stored in the local blockchain of $block~admin$. The public blockchain of $group~head$ might log transactions in certain cases. During the addition of new nodes the $block~admin$ and $group~head$ both store the $access~control$ header and the list of public keys connected with that high processing nodes.
\subsection{Transaction}
\label{leb:tran}
Three types of transactions are available in this proposed architecture namely, $add~node$, $remove~node$ and $data~movement$. 
\subsubsection{Add Node}
The $1^{st}$ network hierarchy has three types of $Add Node$ transactions, such as :  \newline
\textbf{i}. New LP node addition under $block~admin$ supervision. \newline
\textbf{ii}.New $users$ (HP Node) under the $group~head$ supervision. \newline
\textbf{iii}. New $block~admin$ (HP Node) under the $group~head$ supervision.\newline
As shown by ash colour in Fig. \ref{fig:archbc2}, the new addition of LP node with $block~admin$ incorporates the below sub-steps:

The $add~device$ transaction is similar to genesis transaction of bit coin \cite{bitcoin}. The $Add Node$ transaction process of LP generates one public key and one private key. The public key is publicly available and it is inserted in a new blockchain. As suggested by many crypto mathematicians, all the LP and HP nodes of our proposed model use public key for its real world identity \cite{arvind}. But large size of the public key makes it unsuitable for node identity. Hence, the hash value of public key is used as a unique node identity. The private key is kept privately and it is used to sign messages. After the addition of new LP node, sensors and their output interfaces are connected with different LP pins. As the pins connected with LP nodes are unique locally and the previously generated LP node identity is also unique, the concatenated value of both LP nodes' pin and LP nodes' identity are collectively used to generate the sensor/output ID. The new $Add~Node$ operation also creates an access control header for all the connected sensors and outputs. 
     
  Both the new $users$ HP nodes and the new $block~admin$ under the $group~head$ supervision have transactions similar to previous LP node addition under block admin supervision. It follows the bitcoin account creation process along with the access control as mentioned earlier in this work.
The ash colour part of Fig. \ref{fig:archbc2} shows the subsequent process of $add~node$ transaction in $1^{st}$ level network hierarchy. The first step of $add~node$ transaction create a pair of key; a public key and a private key. The private key is kept secret and public key is added to private blockchain of $block~admin$ and public block chain of $group~head$. This indicates that the all HP nodes that belong to the group get informed about the newly added LP node. The same process will generate a node id which is hash value of its public key. In the second step, different sensors and outputs are added with the said LP node. Each sensor and outputs are connected with a unique pin of LP node. Hence, the sensors/outputs id is the concatenated value of local LP pin and the LP node id generated from the hash value of its public key. Say for an example a light sensor is connected with $A0$ pin of a LP node whose node is  $1234567890abcdef$. The id of that light sensor will be $1234567890abcdefA0$. When the sensors and the outputs are fixed for a specific LP node, Algorithm \ref{progcodeiot} can generate the node code which will be downloaded in memory of LP node remotely using Over The Air (OTA) protocol facility \cite{ota}. The ash coloured steps 1 and 2 of Fig. \ref{fig:archbc2} show the add node transaction in the $1^{st}$ level network hierarchy.


\subsubsection{Remove Node}
The $remove~node$ transaction will destroy the said public key from the public key list of its supervisor, and keeping the chain alive in blockchain. The $remove~node$ transaction can occur at $LP~node$, $block~admin$ and $group~Head$. After remove node transaction $LP~node$, $block~admin$ and $group~Head$ respectively can not read or write data from/to $block~admin$, $group~Head$ and other $group~Head$ of canopy network.
\subsubsection{Data Movement}
\label{l:dm}
The proposed architecture consists of two types of data movements which are described below:\\
\textbf{i. LP nodes to Block~Admin:} At 1$^{st}$ level of network hierarchy LP nodes can write sensor data to the $block~admin$ and outputs of LP node can read data from the $block~admin$. It is to be noted that sensor data  writing to $block~admin$ is the reading process of  $block~admin$ from sensors in LP nodes and reading process by outputs is writing process by $block~admin$ to the outputs. For example, say a camera is a sensor and traffic lights are the outputs and both are connected with same or different LP nodes. Camera is sending videos to its $block~admin$ and $block~admin$ calculates the number of cars passing through particular lane where the camera is installed (low processing node to $block~admin$). According to the number of cars, the duration of the traffic light of the said lane is changed dynamically ($block~admin$ to low processing end). 

\begin{figure*}[!htb]
\centering
\includegraphics[scale=0.23]{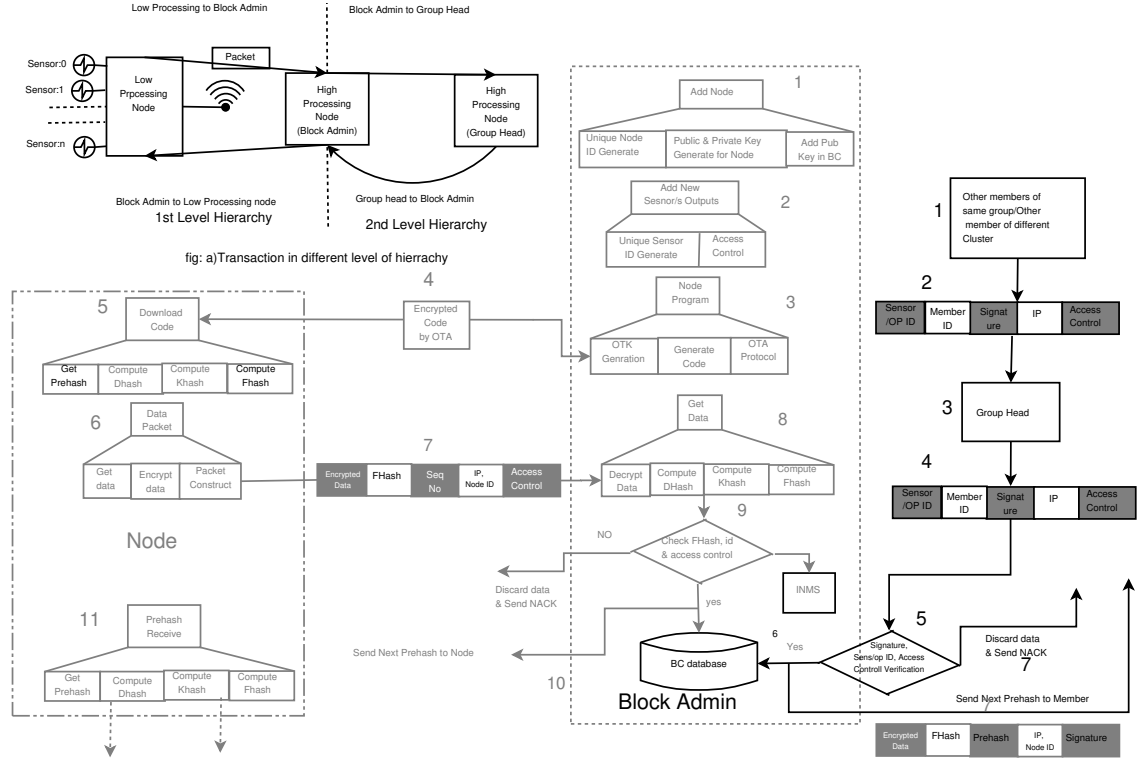}
\vspace{-6pt}
\caption{2$^{nd}$ Level of Hierarchy}
\vspace{-8pt}
\label{fig:archbc2}.
\end{figure*}

As per Fig. \ref{fig:archbc2}, after completion of ash coloured step 1 and 2, the data movement process will start. After generation of node code using OTA protocol, $block~admin$ can periodically configure node code on the specific node. If any attacker tries to modify the node code of any LP node, it will be erased and reconfigured with original code due to the periodic intervention of $block~admin$. The OTA protocol sends the encrypted code to the LP node. During this transmission $block~admin$ generates a one time key (OTK). This OTK is  generated by a hash based on morkel tree on the private key of a LP node and a random number. In the fifth step, the LP nodes are configured and they generate three hash values such as key hash $KHash$, sensor data hash $DHash$ and a final hash $FHash$, where $KHash=hash(OTK+prehash)$.Here $prehash$ is zero for the first data packet and for the other data packets $prehash=Hash(prehash+time+data+error~message)$. $DHash$ is hash of all sensor data and $FHash=Hash(Khash+Dhash)$. At the 6th step, LP nodes encrypt the sensor data and construct a packet with six parameters: encrypted sensor data, FHash, LP node IP, Node ID, packet sequence number and access control header. After that this packet will be sent to $Block~Admin$. At the 8th step the $Block~admin$ will receive the packet followed by a decryption process. The $Block~Admin$ computes $DHash$ from that decrypted data. The $prehash$ is computed by $Block~Admin$ itself, hence this value is already available at the $Block~Admin$ side. $Khash$ will be computed from $OTK$ and $prehash$. The final hash $FHash$ is also computed from $DHash$ and $KHash$. If the computed $FHash$ by $Block~Admin$ matches with the sent $FHash$ by LP, the sensor data will be accepted in the blockchain data base which may be later accessed by other HP nodes of same or different group. If both the $FHash$ match, a new $prehash$ will be generated and sent to the LP node. Else, the data packet will be discarded and $Block~admin$ will ask for a new packet. The whole transaction is shown by ash colour in fig. \ref{fig:archbc2}.\\
\textbf{ii. Block Admin to Group Head:}  After the authenticated permission the HP nodes/members from different group or from the same group can access the data of sensors and output interfaces of $Block~Admin$. As shown in black colour of Fig. \ref{fig:archbc2}, if HP nodes of same group or different group want to access some specific sensors/outputs, they send a packet consisting of sensors/outputs ID, requester member ID, signature of requester, IP of requester member and its access control. The $group~head$ will check the signature and access control. If it matches, access will be granted to the specific sensor data of the LP node, otherwise the situation will be ignored by $group~head$. For example, video captured by a camera need to be written in $block~admin$ to be monitored by some state government or central government agencies through $group~head$ or some other nodes from different groups. Let us say that a traffic signal which is connected as output in some LP nodes changes its timings dynamically according to the vehicle load as calculated from the captured video stored earlier in $block~admin$. In this example, storing video in $block~admin$ as captured by the camera is a writing process to $block~admin$ by camera sensor connected in LP node at first level of network hierarchy. The changes of traffic light timings is a reading process by traffic light outputs  connected in LP node from $block~admin$.
 The two hierarchy transactions are shown in fig. \ref{fig:archbc2}:a.
\subsection{Packet}
In the $1^{st}$ network hierarchy, all the packets travel from LP node to HP node or HP node to LP node. In the $2^{nd}$ network hierarchy, all the packets travel from HP of same or different group to $Block~admin$ or $Block~admin$ to HP of same or different group. The packet configuration is shown in Fig. \ref{fig:local}. The components inside the packet are stated below.
\subsubsection{Symmetric Encryption on Data}
Data collected from $n1$ number of sensors in LP node is encrypted by symmetric key encryption technique. The key for symmetric key encryption algorithm is shared by $block~admin$ HP using Diffie-Hellman algorithm \cite{buchman}. The encryption on sensor data is a mandatory process throughout the sensor data communication from LP to $Block~Admin$ or $Block~Admin$ to HP nodes of same or different group.
\subsubsection{Hash of Data}
LP nodes to $Block~Admin$ and $Block~Admin$ to the HP nodes of the same or different group both use hash on data to serve the data integrity. In the $1^{st}$ network hierarchy, the final $FHash$ depends on $DHash$ where $DHash$ is the hash of sensor data. The same process is also followed in $2^{nd}$ level of network hierarchy. 
The hash in $1^{st}$ level network hierarchy uses light weight hash algorithm.
\subsubsection{Digital Signature}
\label{ds}
LP nodes to $Block~Admin$ and $Block~Admin$ to the HP nodes of same or different group both use signature on data to serve the data availability and authenticity. In the $1^{st}$ network hierarchy, the final $FHash$ depends on $KHash$ where $KHash$ depends on a function of private key. As the private key is secret, no one can generate the appropriate $KHash$ hence signature can not be forged. It is to be noted that in the $1^{st}$ level network hierarchy conventional signature verification algorithm has not been used because signature verification algorithm consumes enormous processor footprint and its throughput is very less which makes it inappropriate for such low processing platform. As all the nodes in the $2^{nd}$ level network hierarchy are high processing platform, hence they use conventional signature verification process during their data transaction.
\subsubsection{Access Control Header (ACH)}
\begin{figure*}[!htb]
\centering
\includegraphics[scale=0.26]{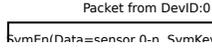}
\vspace{-6pt}
\caption{Local Blockchain}
\vspace{-8pt}
\label{fig:local}
\end{figure*}

The access controller header consists of all the possible read/write permissions in different storage devices. The LPs have different sensors which are generally written in local storage of $block~admin$ or $cloud~storage$ in canopy network. LPs also consist of few outputs which may read data from the $local~storage$ of $block~admin$. For certain applications, different sensors may need different permissions. In Fig. \ref{fig:local} $sensor:0$ of $device:0$ has reading permission  
\subsubsection{Other Data}
The said packet also consists of time stamp, sensor id, device id (LP id).

\section{Auto Code Generator}
The proposed design can generate codes for LP and HP nodes. The auto code generation for LP is  a very crucial step because the $block~admin$ need to aware about the code running in LP to protect LP from malicious code injection. The auto code generation in HP side is also an important issue because the changes  of existing code or creation of new code blockchain technology should be accepted after the distributed consensus of other HP nodes. 
\subsection{Code Generator Algorithm for LP}
While sensors are being added in LP, the $Block~Admin$ can generate LP code automatically by $Auto LP Code$ generator algorithm.
IoT devices are considered as nodes which are denoted by  expression $N_i$, where $i$ is any positive integer from $1$ to $I$. Each node has several sensors and outputs which are denoted by $S_j$ and $O_k$ respectively. $j$ and $k$ are any positive integers which vary from $1$ to $J$ and from $1$ to $K$ respectively. 
The system code $C_i$ of existing IoT node has 6 sections as given below:
\subsubsection{Static Node libraries}
The IoT node libraries $SNL$ is needed for crypto applications and other network components of IoT device. The $SNL$ is not dependent on the type of IoT node. Like WiFi client libraries, MD5Builder and other crypto library may work with almost all the type of IoT nodes.
\subsubsection{Dynamic Node Libraries}
The dynamic node libraries $DNL$ depends on board type, once the board type is changed the dynamic node library expression is changed. Hence the $DNL$ can be stated by Eq. \ref{equ:DNL} where $DNL_i$ is node function ($f_{n_i}$) of $N_i$ node.
\vspace{-10pt} 
\begin{equation}
\label{equ:DNL}
DNL_i=f_n(N_i);
\end{equation}
For example, the web server library for $esp32$ and $esp8266$ use different packages.

\makeatletter
\def\BState{\State\hskip-\ALG@thistlm}
\makeatother
\alglanguage{pseudocode}
\begin{algorithm}[]
\small
\caption{Auto LP Code Algorithm Structure}
\begin{algorithmic}[1]

\Procedure{$\mathbf{Input}$}{i,$\bigcup_{j=1}^J p_j, t_j$, $\bigcup_{k=1}^K (p_k, t_k )$} 
\State SNL
\State $DNL_i$

    \For {$j = 1 \to J$}
    \State $f_{DSL}(f_{sj}(p_j, t_j ))$
    \EndFor
       \For {$j = 1 \to J$}
    \State $f_{po}(f_{sj}(p_j, t_j ))$
    \EndFor
           \For {$k= 1 \to K$}
    \State $f_{po}(f_{sj}(p_k, t_k ))$
    \EndFor
    \For {$j = 1 \to J$}
    \State $f_{su}( \bigcup_{j=1}^J f_{sj}(p_j, t_j )$;
    \EndFor
    \State $SUN$
               \For {$j= 1 \to J$}
    \State $f_{sj}(p_j, t_j )$;
    \EndFor
               \For {$k= 1 \to K$}
    \State $ f_{ok}(p_k, t_k )$;
    \EndFor
    
                   \For {$j= 1 \to J$}
    \State $f_{sj}(p_j, t_j )$;
    \EndFor
   
                  \For {$k= 1 \to K$}
    \State $f_{ok}(p_k, t_k )$;
    \EndFor
\EndProcedure
\end{algorithmic}
\label{progcodeiot}
\end{algorithm}
\subsubsection{Dynamic Libraries for Sensors}
Dynamic libraries of sensors are denoted by $DSL_{S_j}$ which depends on sensors $f_{s_j}$ connected with nodes. The sensor function $f_{s_j}$ depends on the type of sensor $t_j$ and the connected pin $p_j$. For $J$ number of sensors, $DSL_{S_j}$ can be expressed by Eq. \ref{equ:dsl}. It is to be noted that $DSL_{S_j}$ is independent of node $N_i$.
\begin{equation}
\label{equ:dsl}
DSL_{S_j}=f_{DSL}(\bigcup_{j=1}^J f_{sj}(p_j, t_j ));
 \end{equation}
 \subsubsection{Pins and Objects}
 Pins and objects (if needed) $PO_{S_jO_k}$ of sensors and outputs are sensor function $f_{sj}$ and output function $f_{ok}$ which depend on $p_j$, $t_j$ and $p_k$, $t_k$ respectively. The equation of $PO_{S_jO_k}$ is stated in Eq. \ref{equ:po}.
 \begin{equation}
\label{equ:po}
PO_{S_jO_k}=f_{po}( \bigcup_{j=1}^J f_{sj}(p_j, t_j ), \bigcup_{k=1}^K f_{ok}(p_k, t_k ));
\end{equation}
 
\subsubsection{Set up function} 
The setup function consists of two parts; one part is static for boards and network application which is denoted by $SUN$ and other part $SU_{sj}$ is for sensors. The expression of $SU$ can be stated by Eq. \ref{equ:su}.
 \begin{equation}
\label{equ:su}
SU=f_{su}( \bigcup_{j=1}^J f_{sj}(p_j, t_j ), + SUN);
\end{equation}
\subsubsection{Data packet}
Data packet $DP_{S_j}$ generation for sensors $S_j$ and receiving packet $RP_{O_k}$ construction for outputs $O_k$ are very important for IoT platforms. The expression of data packet $DP$ can be stated by Eq. \ref{equ:dp}.
 \begin{equation}
\label{equ:dp}
DP_{S_jO_k}=f_{dp}( \bigcup_{j=1}^J f_{sj}(p_j, t_j ), \bigcup_{k=1}^K f_{ok}(p_k, t_k ));
\end{equation} 
 \subsubsection{Delay}
 The delay $D_{so}$ for sensors and outputs are given by the users. Depending on the application and sensor data, the delay may be varied. The expression of delay can be expressed by Eq. \ref{equ:delay}.
 \begin{equation}
\label{equ:delay}
D_{S_jO_k}=f_{dp}( \bigcup_{j=1}^J f_{sj}(p_j, t_j ), \bigcup_{k=1}^K f_{ok}(p_k, t_k ));
\end{equation}  
 \par The final system code $c_i$ of $N_i$ node will be the function ($f_c$) of all the aforementioned six parameters. The expression of final code can be stated by Eq. \ref{equ:finalcode}.
 \vspace{-10pt} 
\begin{equation}
\small
\label{equ:finalcode}
\begin{split}
c_i&=f_c(SNL,DNL_i, f_{DSL}(\bigcup_{j=1}^J f_{sj}(p_j, t_j )),\\
& f_{po}( \bigcup_{j=1}^J f_{sj}(p_j, t_j ), \bigcup_{k=1}^K f_{ok}(p_k, t_k )), f_{su}( \bigcup_{j=1}^J f_{sj}(p_j, t_j ), \\
&+SUN, f_{dp}( \bigcup_{j=1}^J f_{sj}(p_j, t_j ), \bigcup_{k=1}^K f_{ok}(p_k, t_k )), \\
&f_{dp}( \bigcup_{j=1}^J f_{sj}(p_j, t_j ), \bigcup_{k=1}^K f_{ok}(p_k, t_k )));
\end{split}
\end{equation}

\subsection{Code Generator Algorithm for HP}
\label{l:codehp}
In every $Block~Admin$, there should be a code to receive sensor data or to send data to outputs. For each sensor and output there will a separate HP code running in $Block~Admin$. If there are $n1$ number of sensors and $n2$ number of outputs are connected in each LP node and the total number of LP nodes under a $Block~Admin$ is $l$, then the number of HP codes to receive sensor data and to send output data is $l\times(n1+n2)$. The different parts of the HP code are stated below sequentially.
\begin{enumerate}
\item This part of the code receives all the parameters sent by LP node. The details of all these parameters are discussed in Section \ref{leb:tran}. 
\item $Block~Admin$ generates the time of the data receipt. It is to be noted that the time stamp can be generated in LP node itself and can be sent as a parameter. The implementation experience suggests that the generation of timestamp in LP node is an extra overhead and it does not matter if we consider the $Block~Admin$'s receiving time as a time stamp of the data. Additionally reducing the number of parameter reduces the Internet traffic.
\item Extract private key using the LP node id to generate the One Time Key (OTK).
\item Generate $KHash$ from $OTK$ and $prehash$. Generate $DHash$ after decrypting the encrypted by its symmetric key. 
\item Check $FHash$ and decide for the acceptance of data.
\item Generate $prehash$ for next data.  
\end{enumerate}
%
\begin{table}[!htb]
\caption{Features Table} 
\centering  
\resizebox{9cm}{!}{%
    \begin{tabular}{|c|c|}
        \hline
Security Issues	& Proposed Solution\\\hline
Confidentiality&	Symmetric Key encryption used for all transaction\\\hline
Integrity &	All transaction consist hash.\\\hline
Availability & 	The access controller header of local and canopy network ensuring that \\ 
& all systems services are available, when requested by an authorized user.\\\hline
Authentication&	It is achieved by using access header and shared keys.\\\hline
Nonrepudiation	&All local and canopy transactions are signed by the \\ 
&transaction generator to achieve non-repudiation. Additionally, Hence \\
&neither requester nor requestee can deny their complicity in a transaction\\\hline
    \end{tabular}
}
\label{table:fet} 
\end{table}
\begin{table*}[!htb]
\vspace{-5pt}
\caption{Threat \& Defence} 
\centering  
\resizebox{15cm}{!}{%
    \begin{tabular}{|c|c|c|c|c|}
        \hline
Name of 	& Definition & Defence & Tolerance &Sustenance\\
Attack        & ~	     & ~     & &\\\hline
\shortstack{Denial of Service\\ (DoS)\\~\\~\\~\\~}  &	\shortstack{Attacker attacks local canopy network node with\\   a huge number of transactions to flood out the \\   node such that it cannot commit any genuine tr-\\ -ansactions from other nodes.}		 &\shortstack{Nodes in canopy and local network could not send a (see\ref{l:dm})\\ transaction to its group members unless its key finds a match in \\ their keylist. As stated in sec. \ref{sec:time} each nodes in canopy and\\ local network have a threshold of transaction rate. If  it exceeds\\  threshold, a penalty will be charged.} & \shortstack{Very High\\~\\~\\~\\~} &\shortstack{Unlikely\\~\\~\\~\\~}\\\hline

\shortstack{ Distributed  Denial\\ of Service (DDoS)\\~\\~\\~}  & \shortstack{The distributed version of
the DoS attack, where \\   multiple local or canopy  nodes  are compromised.\\~\\~}			 &	\shortstack{ The nodes in canopy and local network are not diretcly accessible.\\ The key list restricted unautheticated nodes. The nodes in under \\ local or canopy  network can read or write data of other nodes if\\  key has been shared (see \ref{l:dm} \& \ref{leb:tran})}	 & \shortstack{High\\~\\~\\~}&\shortstack{Unlikely\\~\\~\\~\\~}\\\hline
\shortstack{Injection\\node}     &	\shortstack{Attacker injects fake nodes into network to\\  get access to private data}		 &\shortstack{The injected node is isolated as local networks require shared\\  key, which needs approval from the $block~admin$ (see \ref{smrt:blk})}		 & \shortstack{Moderate\\~} &\shortstack{Possible\\~}\\\hline

\shortstack{Appending\\~\\~}     &	\shortstack{Attacker attacks canopy and local
 network gene- \\-rates fake transactions to create false reputation}& \shortstack{Block Admin \& Group head can detect malicious nodes\\ during the signature verification step (see \ref{sec:contact} and \ref{ds}.)}		 & Very High&Unlikely\\\hline
 
\shortstack{Consensus Period\\Attack }   &	\shortstack{Attacker attacks canopy and
 generates fake\\ transactions to create false reputation\\~}		 &	\shortstack{Transaction will be valid, if at least half the  number\\ of nodes will sign. This possibilities is very low.}	 & \shortstack{High\\~\\~} &\shortstack{Unlikely\\~\\~}\\\hline 

\shortstack{Dropping\\Attack\\~\\~ }   &	\shortstack{The Group Head does not response transactions\\ to or from its group members, hence nodes are\\  isolate them from the canopy}		 &	\shortstack{A group member can alter its group head if it\\ notices that its transactions process is failed.\\~}	& \shortstack{High\\~\\~} &\shortstack{Unlikely\\~\\~}\\\hline 

\shortstack{Linking\\Attack\\~\\~\\~\\~ }   &	\shortstack{Attacker as service provider or cloud storage adds\\ multiple transactions in the BC with the same identity\\  to get the real world identity of an anonymous node\\~}	& \shortstack{The nodes in canopy use unique public key for each\\ transaction. This restricts the attacker from linking\\ the data of multiple nodes of the same user.}	&	\shortstack{Very High\\~\\~\\~} &\shortstack{Unlikely\\~\\~\\~}\\\hline 

\shortstack{Routing\\Attack\\~\\~ }   &	\shortstack{Transactions propagating through\\ the local or canopy network and\\ tampering with them before passing}		 & \shortstack{It uses hashing in local and canopy network(see \ref{l:codehp})\\~\\~}		
&	\shortstack{Very High\\~\\~} &\shortstack{Unlikely\\~\\~}\\\hline
    \end{tabular}
}

\label{table:attack} 
\end{table*}

\vspace{-10pt} 
\section{Smart Contract Framework}
\label{sec:contact}
As shown in Fig. \ref{fig:local}, every node will have access control header (ACH) which defines reading and writing permissions of every sensor/output connected with LP nodes. Before granting any transaction in any network hierarchy of this proposed model signature, access control header and POW (optional in private BC) verifications are needed. Along with this, also a parameter in percentage format named as probability of misbehaviour (PM) is calculated. According to the percentage of PM, the specific node will be charged a penalty that is, the said node might be blocked by the respondent for a certain period of time. To measure the PM, we have incorporated the below three inconsistency parameters.
\vspace{-10pt} 
\subsection{Time Inconsistency}
\label{sec:time}
\subsubsection{Minimum Frequency}
The minimum frequency ($minFreq$) of sensor/output is the minimum permissible time interval between two successive transaction requests. In runtime, the smart contract algorithm (SMA) calculates the time interval between two successive transaction requests; if it is less than $minFreq$ the latter transaction request will be treated as a frequent request. 
\subsubsection{The Number of Frequent Transaction Requests(NoFTR)} 
The access control header will store a cut-off for each sensor/output. If the number of frequent transaction requests (NoFTR) is larger than or equal to the cut-off, the SMA detects a misbehaviour occurrence.
\subsubsection{Time of last Transaction (ToLT)}: the time of the last access
transaction from the sensor.
\subsubsection{Time Stamp}
This is the time when the sensor data from LP node is received in $Block~Admin$. The details of the timestamp is discussed in section \ref{l:codehp}. To check the time inconsistency the following steps are considered:
\begin{itemize}
\item Access control header will be checked against the sent sensor id.
\item The subtracted value of ToLT from $timestamp$ is the run time interval between two successive transaction requests. If it is less that $minFreq$, $NoFTR$ will be increased by 1. If $NoFTR$ reaches a certain threshold, the said node will be considered compromised under the time inconsistency issue.
\end{itemize} 
\subsection{Data Inconsistency}
The proposed application uses two types of data inconsistency detection. The first one is data inconsistency detection based on Dempster-Shafer Theory of Evidence (centralised detection) and the second stage detects data inconsistency based on mean distance on sensor data from neighbour sensor nodes (distributed detection).
\subsubsection{Dempster-Shafer Theory of Evidence on Centralized Sensor node}
	Dempster-Shafer evidence theory \cite{dempster},\cite{shafer} is used to address the uncertainty in statistical conclusions. Statistical inference includes all the possible states of a system which are generally termed as hypotheses. These hypotheses are then assigned mass functions based on which some deductions are made. Dempster-Shafer evidence theory helps in data fusion of different sensor data by applying the said mass functions. Usually, a single mass function is used to represent a single data source. But to reach a conclusive decision based on sensor data fusion, it is necessary to have a cumulative value. This problem is addressed by Dempster's combination rule, which is given by:
	\begin{equation}
	\label{eq:1}
	m_1\oplus m_2 (Z) =\frac{\sum_{X \cap Y=Z\neq \phi} m_1(X)m_2(Y)}{1-K}
	\end{equation}
	Here, $K = \sum_{X \cap Y=\phi} m_1(X)m_2(Y)$ and $X,Y,Z \subseteq \Theta$. $\oplus$ is the orthogonal or direct sum. So, $m_1\oplus m_2 (Z)$ is the combined evidence of two different mass assignments and $\phi$ is the null set. The numerator in (\ref{eq:1}) covers all the possibilities whose intersection is $X \cap Y=Z$. To normalise this value, it is divided by $1-K$ which represents all the combined values that produces a null set. Dempster's rule of combination is iteratively applied on all the information sources to produce the final result.
	
	In this work, it is assumed that the data is normally distributed. One of the primary reasons for this assumption is that the data set is sufficiently large which is an inherent property of a data set to have normal distribution. Now that the data distribution is taken care of, a very important part is designing the mass function. For determining the state of a sensor to be either faulty or normal at a primary level, the following mass function is designed:
	\begin{equation}
	m_\lambda(\delta)=\frac{\rho_\lambda(\delta)}{\sum_\delta \rho_\lambda(\delta)}
	\label{ma} 
	\end{equation}
	where, $\rho_\lambda(\delta)$ is the probability density function and is given as:
	\begin{equation}
	\rho_\lambda(\delta) = \frac{1}{\sqrt{2\pi \sigma_{\delta\lambda}^2}}\exp^-{\frac{(x_{t_\lambda}-\mu_{\delta\lambda})^2}{2\sigma_{\delta\lambda}^2}}
	\label{pdf} 
	\end{equation}
	The notations as given in Eq.~\ref{ma} are as follows:
	\begin{enumerate}
		\item [$\bullet$] $\sigma_{\delta\lambda}$ : Standard deviation of the training set for the feature $\lambda$ in class $\delta$
		\item [$\bullet$] $\mu_{\delta\lambda}$ : Expectation of the training set for the feature $\lambda$ in class $\delta$
		\item [$\bullet$] $x_{t_\lambda}$ : Value of feature $\lambda$ of the test vector $x_t$ 
	\end{enumerate}
After $m_\lambda(\delta)$ is calculated, the mass functions are combined according to Eq.~\ref{ma}. Then the conclusion is based on the combination rule which states that if the combined mass assignment for normal sensor is greater than the combined mass assignment for faulty sensor, then the sensor is normal, else the said sensor will be considered compromised in-terms of data inconsistency under centralized consensus. The details of this work is reported in our previous work \cite{ng}.
\subsubsection{Distance Neighbour Mean on Distributed Sensor Node} 
It calculates data inconsistency based on mean data distance form a given node to its neighbour node. If that Distance Neighbour Mean (DNM) exceeds certain given threshold, the sensor node will be considered to be faulty. The GPS sensor provides the latitude and longitude of the LP node. If $x_i$ and $y_i$ are the latitude and longitude of the given $i^{th}$ LP node, we configure a virtual square window of $s$ arm around the coordinate $x_i$ and $y_i$.  The four coordinates of the square will be ($x_i-s/2$, $y_i+s/2$), ($x_i+s/2$, $y_i+s/2$), ($x_i-s/2$, $y_i-s/2$) and ($x_i+s/2$, $y_i-s/2$).  We will calculate mean of distance from $i^{th}$ LP node to all the node coordinates under this virtual square. Say, we have $n1$ temperature sensors connected with $n1$ LP nodes inside the said square. The DNM of temperature sensor of $i^{th}$ node will be calculated by Eq. \ref{equ:dnm} where $t$ is the temperature of the $i^{th}$ LP node and $t_j$ is the combined temperature of the surrounding LP nodes, where $j$ varies from $1$ to $n1-1$.
	\begin{equation}
	\label{equ:dnm}
	DNM_i=\sum^{n1-1}_{j=1}|t-t_j|
	\end{equation}
If the $DNM$ exceeds a given threshold, the temperature sensor of $i^{th}$ LP node will be considered compromised in terms of data inconsistency under distributed consensus.
\subsection{Network Inconsistency}
As discussed in Section \ref{l:dm}, the decrypted sensor data will be accepted in BC ledger when $FHash$ is matched. This process covers all the cryptographic issues as discussed in Table \ref{table:fet}. $Fhash$ will not match if any crypto process is compromised by an intended attacker putting inconsistency in the network part. The malfunction might also occur due to technology failure. Both these situations are considered as network inconsistency. Any $FHash$ mismatching indicates that the node is compromised due to network inconsistency. $FHash$ matching is considered as database hit, else it is a database miss. We compute a ratio of database hit and database miss (database hit+database miss) to measure the reliability of the given LP node. 
The threat models of the proposed article is discussed in Table \ref{table:attack}.
\section{Results and Implementation}
\label{sec:rai}
The proposed architecture uses Node MCU as LP which is connected with five sensors such as temperature, humidity, pressure, camera, IR and light intensity. DHT11 sensor is used for humidity and temperature measurement. BMP 180, $flying~fish$ and LDR are used as pressure sensor, IR and Light Detecting Resistance (LDR) respectively. The $ArduCAM Mini module$ is used as camera. Conventional 8GB RAM computer with and i7 processor is used as $block~admin$. The private blockchain, access control header, public key list of LP in 1$^{st}$ level hierarchy are stored in mysql database. The APIs  of local network are written in PHP to extract all the required informations from the transaction packets. The Canopy network is design in Etherium platform. 
\par Table \ref{table:fet} ensures that the five primitives of cryptography  such as confidentiality, authenticity, integrity, availability  and repudiation are managed. In Table \ref{table:comp}, the reported architecture is compared with the existing bit coin architecture \cite{bitcoin}, \cite{zhang}, \cite{afar} and a recent bench mark implementation \cite{alidori}. The article \cite{alidori} is the most popular and the most cited blockchain based IoT implementation in existing literature. Our work has found seven major changes over the \cite{alidori} and other existing literature as shown in of Table \ref{table:comp} marked in gray. The 3$^{rd}$ row of Table \ref{table:comp} states about the $Miner~Joining~Overhead$ which is the most profound issue of blockchain architecture. 
\par In existing blockchain of bitcoin \cite{bitcoin}, \citep{afar}, \cite{zhang} and Ali Dori et al. \cite{alidori}, the newly joined miner needs to download all the blockchain of the previous transactions. The proposed work also downloads all the previous transactions in current blockchain and additionally it includes the public key of the newly joined user into a separate private blockchain. As per section \ref{sec:lpinit}, it is to be noted that for newly joined LP it does not have such downloading overheads. This blockchain of the public keys of $block~admin$ stores all the public keys of the authorized HP users and LP. Any inclusion or exclusion of HP/LP nodes in this blockchain is tracked cryptographically. The adoption of the BC technology to store public keys of devices avoids situations where malicious devices can steal network password and connect with $block~admin$. The absence of public key of malicious device in public key blockchain blocks that device from transaction of any packets. This scheme prevents device injection attack. appending attack, DoS and DDoS attack in local and canopy networks as stated in Table \ref{table:attack}. The $group~head$ and other HP nodes in canopy network always validate each new transaction that it receives from
other $group~head$ before to appending it to the public BC. The validation process consists
signature verification process. It is to be noted that each $group~head$ and other HP nodes in canopy use a pre-shared public key for generating transactions and it is assumed that these public keys
are shared to all other $group~head$ and other HP nodes in canopy which makes the architecture safe from appending attack. 
\begin{table*}[!htb]
\vspace{-10pt}
\caption{Comparison of Proposed Blockchains with existing Blockchains} 
\centering  
\resizebox{12cm}{!}{%
    \begin{tabular}{|c|c|c|c|c|c|c|c|c|c|}
        \hline
Parameter 	& Bit Coin & Local BC & Public BC& Proposed  & Proposed &&&Local BC & Public BC\\
~	& \cite{bitcoin} & \cite{alidori} & \cite{alidori}&  Local BC &  Public BC & \cite{zhang}&\cite{afar}&\cite{rourab} & \cite{rourab} \\\hline
Mining& PoW & None & None &  None &  None&PoW&PoW&  None&  None\\\hline
BC Scope	& Public & Private & Private &  Private  &  Private &Public&Public&  Private &Private\\\hline
\rowcolor[gray]{0.7}User 	& 	download 	& ~download all &  download all & Block admin down &  ~download all &Download &Download&Block admin dow&download all\\
\rowcolor[gray]{0.7}Joining  & 	all blocks 	& blocks in BC & blocks in BC &  load all  blocks in  &  blocks in BC, &all the blocks&all the blocks&load all  blocks in &blocks in BC\\
\rowcolor[gray]{0.7}Overhead		& in BC		& ~ & ~ & BC, add public key&  add public key &in BC&& BC, add public key&  add public key\\\hline
BC Control	& 	None	& Owner & None &  Owner&  None &None&None& Owner & None\\\hline
Double	& Not Possible & NA & NA &  NA &  NA &NA&NA&  NA &  NA\\
 Spending	& 		& ~ & ~ &  ~ &  ~ &&&&\\\hline
Transaction	& Broadcast		& Unicast & Unicast &  Unicast &  Unicast &Unicast&Broadcast/ & Unicast & Unicast\\
Type	& 		& ~ & /Multicast &  &  /multicast & /multicast&Multicast& ~ & /Multicast\\\hline
Transaction	& input, coin     & Block-no., 	  & Output, &  Block-no., &  Output, &Output&Output& Block-no., 	  & Output, \\
parameter 	&  output 		& hash data, PK  & PKs    & hash data, PK  &  PKs &Result&Policy,& hash data, PK  &PKs\\
~			& 			& time, output,    & & time, output, &  ~ &Penalty&lock,& time, output,&\\
			& 			&  policy rules.   & ~ &  policy rules.&  ~ &&Unlock&  policy rules&\\\hline
\rowcolor[gray]{0.7}Block Header	& Hash Puzzle  & Polices 		& Policies &  Access header &  Access header &Access header&Locking script&  Access header &  Access header\\\hline
\rowcolor[gray]{0.7}Encryption	& Public Key 	 & Not 		 & Public key 	    &  Public key &  Public key &Information&Public Key& Public key 	    &  Public key\\
\rowcolor[gray]{0.7}process		& Cryptography  &  studied 	& Symmetric Key &  Symmetric Key &  Symmetric Key &Unavailable&	& Symmetric Key &  Symmetric Key \\\hline
Forking	& Not Allowed & Allowed & Allowed  &  Allowed  &  Allowed  &Allowed&Allowed&  Allowed  &  Allowed \\\hline
Reward	& Coins & Nothing & Not defined &  Nothing &  Nothing &Nothing&nothing&Nothing&nothing\\\hline
Pool		& Allowed& cannot be & cannot be  &  cannot be &  cannot be &Not&Not& cannot be  &  cannot be\\
Mining	& 		& defined & defined &  defined &  defined & Defined&Defined & Defined&Defined\\\hline
\rowcolor[gray]{0.7}Malicious	& Allowed	& possible & not possible &  not possible &  allowed &Possible&Possible&  not possible &  allowed\\
\rowcolor[gray]{0.7}User	& 		& ~ & ~ &  ~ &  ~ &&&&\\\hline
Miner	& 	Self		& Owner& Node in group &  Owner &  Node in group &Self&Self& Owner& Node in group\\
Selection	& 	Selection	& choice & Choice&  Choice &  Choice &Selection&Selection& choice & Choice\\\hline

\rowcolor[gray]{0.7}Misbehaviour	& No & No   & No  & Yes &  No & Yes &Yes& No  & No\\
\rowcolor[gray]{0.7}Study& 	& 		& ~ & (ML based) &   &  ~ &&  ~ &\\\hline

\rowcolor[gray]{0.7}LP Code	& Manual & Manual   & Manual  & Automatic &  Manual & Manual &Manual& Manual &Manual\\
\rowcolor[gray]{0.7}Generation& 	& 		& ~ & ~ &  ~ &  ~ &&  ~ &\\\hline

\rowcolor[gray]{0.7}HP Code	& Manual & Manual   & Manual  & Automatic &  Manual & Manual &Manual& Manual &Manual\\
\rowcolor[gray]{0.7}Generation& 	& 		& ~ & ~ &  ~ &  ~ &&  ~ &\\\hline

    \end{tabular}
}
\label{table:comp} 
\end{table*}
The 8$^{th}$ row of Table \ref{table:comp} shows that block header architecture is different for all existing architecture. The 9$^{th}$  row of table  \ref{table:comp} shows that our architecture uses encryption in local and canopy network where as \cite{alidori} used encryption only the overlay network. Other literatures such that \cite{bitcoin},\cite{afar} and \cite{zhang} does not have such types of local and overlay network. Hash puzzle and policy header are used article \cite{bitcoin} and \cite{alidori} respectively where as we use an access control process on block header. The 13$^{th}$ row of Table \ref{table:comp} shows that malicious user can participate \cite{bitcoin}, \cite{afar}, \cite{zhang} and local BC of \cite{alidori} which is considerably tough in our local BC because of the schemes stated in section \ref{sec:contact}. The rejection of malicious users as stated in section \ref{sec:contact} ensures Denial of Service (DoS) attack is not possible in proposed article because canopy and local network could not send a transactions to their group members
unless they find a match in their key-list. The 15$^{th}$ row shows that misbehaviour of nodes has not studied in \cite{alidori} and \cite{bitcoin} where as \cite{zhang},\cite{afar} and our proposal put significant focus on this relevant issue. To the best of our knowledge our proposed work is the first BC based IoT architecture which can generate codes for HP and LP nodes as stated 16$^{th}$ and 17$^{th}$ rows of table \ref{table:comp}.   
\section{Conclusions}
\label{sec:con}
Standard security protocols are not suitable for IoT application due to its immense time-space requirements. This proposed article modified the architecture of conventional blockchain technology to adopt it in IoT application. The automated code generation for $block~admin$ and IoT nodes make the platform user convenient and unique. This architecture preserve authenticity, confidentiality,  availability, integrity and non-repudiation. Several standard attacks such as stated in Table\ref{table:attack}  can be prevented by this ML based smart access controlled blockchain platform used in IoT application. The comparison with existing literature establishes that the proposed architecture is better in terms security issues.



%
\bibliographystyle{unsrt}  
\bibliography{bare_jrnl}

\end{document}